\documentclass[prl,twocolumn,superscriptaddress]{revtex4}
\usepackage{amstext}
\usepackage{amsmath}
\usepackage{amssymb}
\usepackage{graphicx}
\usepackage{latexsym}
\usepackage{bm}

\input{tcilatex}
\begin{document}

\title{Quantum criticality in a generalized Dicke model}
\author{Yong Li}
\affiliation{Institute of Theoretical Physics \& Interdisciplinary Center of Theoretical
Studies, Chinese Academy of Sciences, Beijing, 100080, China}
\affiliation{Department of Physics \& Center of Theoretical and Computational Physics,
The University of Hong Kong, Pokfulam Road, Hong Kong, China}
\author{Z. D. Wang}
\affiliation{Department of Physics \& Center of Theoretical and Computational Physics,
The University of Hong Kong, Pokfulam Road, Hong Kong, China}
\author{C. P. Sun}
\affiliation{Institute of Theoretical Physics \& Interdisciplinary Center of Theoretical
Studies, Chinese Academy of Sciences, Beijing, 100080, China}
\date{\today }

\begin{abstract}
We employ a generalized Dicke model to study theoretically the quantum
criticality of an extended two-level atomic ensemble interacting with a
single-mode quantized light field. Effective Hamiltonians are derived and
digonalized to investigate numerically their eigenfrequencies for different
quantum phases in the system. Based on the analysis of the eigenfrequencies,
an intriguing quantum phase transition from a normal phase to a
super-radiant phase is revealed clearly, which is quite different from that
observed with a standard Dicke model. 
\end{abstract}

\maketitle

\section{Introduction}

Quantum phase transition (QPT) and quantum critical phenomena, which are
induced by the change of parameters and are accompanied by a dramatic change
of physical properties, occur at zero temperature in many-body quantum
systems \cite{sachev}. Usually, a QPT may emerge in the parameter region
where there is the energy level crossing or the symmetry-breaking. QPTs have
been mainly studied 
in connection with correlated electron and spin systems in condensed matter
physics \cite{sachev}. 
Very recently, it has also been paid much attention in the light-atoms
interacting systems \cite{Reslen05,emary}, which enables us to understand
the transition from radiation to super-radiation from a different viewpoint.

The systems of atomic ensembles interacting with optical fields have been
studied both experimentally and theoretically, e.g., the electromagnetic
induced transparency \cite{Harris} and the quantum storage of photon states 
\cite{lukin,sun-li-liu-prl}. The thermal phase transition phenomena \cite%
{Rzazewski} have been studied in the Dicke model \cite{Dicke54} (that is, a
two-level atomic ensemble coupling with optical field) or generalized Dicke
models \cite{Hioe}. In particular, the QPT in a radiation-matter interacting
system was recently explored based on the Dicke model, but merely with the
single-mode Dicke model \cite{Reslen05,emary,emary2,Vidal}. When the
coupling parameter $\lambda $ varies from that less than the critical value $%
\lambda _{c}$ to that larger than $\lambda _{c}$, the system goes from the
normal phase to the super-radiant one in the presence of the
symmetry-breaking. As the precursors of the QPT, the onset of chaos \cite%
{emary} and the entanglement properties \cite{emary2} 
were studied in detail. However, it is noticed that these studies 
focused only on atomic ensembles with small dimensions compared with the
optical wavelength, in which the dipole approximation can be used \cite%
{Reslen05,emary,emary2}. In this special case, the light-atoms interaction
is irrelevant to the spatial positions of atoms. But, generally speaking, a
realistic atomic ensemble may extend in a large scale so that light-atoms
interaction is spatially dependent \cite{Dicke54,scully06}.

In this paper, an exotic QPT phenomenon is investigated theoretically by
developing the Dicke model for a more general case beyond the dipole
approximation. We find that a kind of quantum critical phenomenon also
occurs in this extended atomic ensemble, in which each atom interacts with a
single mode quantized light field; but the quantum criticality is quite
different from that deduced from the spatially independent Dicke model \cite%
{emary}. 
In the present study, a normal phase and four possible super-radiant phases
are found, with only one of the four exhibiting the same critical point as
that in the normal phase. Remarkably, it is shown that the ground-state
energy in the above-mentioned superradiant phase, connects continuously to
the normal phase one at the critical point, but its second drivative does
not.

\section{A generalized Dicke model for an extended atomic ensemble}

Let us consider an extended ensemble with $N$ identical two-level atoms
interacting with a single-mode quantized light field. Here, the spatial
dimension of the atomic ensemble is much larger than the optical wavelength
of the field. This radiation-matter system is usually described by a
generalized Dicke model \cite{Dicke54} with the Hamiltonian $H=H_{0}+H_{I}$, 
\begin{eqnarray}
H_{0} &=&\omega a^{\dagger }a+\omega _{0}\sum_{j=1}^{N}\sigma _{\mathrm{ee}%
}^{(j)},  \notag \\
H_{I} &=&\frac{\lambda }{\sqrt{N}}\sum_{j=1}^{N}(a^{\dagger }\mathrm{e}%
^{-ikr_{j}}+a\mathrm{e}^{ikr_{j}})(\sigma _{\mathrm{eg}}^{(j)}+\sigma _{%
\mathrm{ge}}^{(j)}).  \label{001}
\end{eqnarray}%
Here, $\hbar =1$, $k$ is the wave vector of the quantized light field and $%
r_{j}$ is the position of $j$th atom; $\sigma _{\mathrm{ee}}^{(j)}$ is the
population operator of the $j$ atom; $\sigma _{\mathrm{eg}}^{(j)}$ is the
flip operators between the excited state $\left\vert e\right\rangle $ and
ground state $\left\vert g\right\rangle $ of the $j$ atom with the same
energy differences $\omega _{0}$ for all the atoms; $a$ ($a^{\dagger }$)\ is
the annihilation (creation) operator of the quantized light field, with $%
\lambda $ the relative coupling parameter. For simplicity, the ensemble is
assumed to be one-dimensional with its direction along the wave vector. 

Different from a standard Dicke model for small-dimension atomic ensembles 
\cite{emary}, the spatial-dependent factors $\exp (\pm ikr_{j})$ are taken
into account seriously though the momentum of the center of mass can be
neglected. It is also remarked that the terms connected to the
nonrotating-wave scenario are still kept in Hamiltonian (\ref{001}); in
fact, if the rotating-wave approximation were used, the factors $\exp (\pm
ikr_{j})$ would be absorbed into $\sigma _{\mathrm{eg}}^{(j)}$ and $\sigma _{%
\mathrm{ge}}^{(j)}$ \cite{Fleischhauer05,Li05}.

\section{Normal phase}

We can first introduce the following collective operators \cite%
{sun-li-liu-prl}: 
\begin{eqnarray}
B^{\dagger } &=&\frac{1}{\sqrt{N}}\sum_{j=1}^{N}\sigma _{\mathrm{eg}}^{(j)}%
\mathrm{e}^{ikr_{j}},\text{ }  \notag \\
C^{\dagger } &=&\frac{1}{\sqrt{N}}\sum_{j=1}^{N}\sigma _{\mathrm{eg}}^{(j)}%
\mathrm{e}^{-ikr_{j}}.  \label{collective}
\end{eqnarray}%
It is obvious that in the limit of large $N$ with a small number of
excitations (referred to as the normal phase), namely, the excitation
numbers in states $\left\vert e\right\rangle $\ are much less than $N$, the
above two operators approximately satisfy the independent bosonic
commutation relations%
\begin{equation}
\left[ B,B^{\dagger }\right] \approx \left[ C,C^{\dagger }\right] \approx 1,%
\text{ \ }\left[ B,C^{\dagger }\right] \approx 0
\end{equation}%
in the present extended ensemble, and can approximately be re-expressed as
the two independent Bose operators $b^{\dagger }$ and $c^{\dagger }$%
\begin{equation}
\left[ b,b^{\dagger }\right] =\left[ c,c^{\dagger }\right] =1,\text{ \ }%
\left[ b,c^{\dagger }\right] =0.
\end{equation}%
In the present normal phase case, the original radiation-matter system
described by Eq. (\ref{001}) is approximated as a coupling three-mode
bosonic system with the \textquotedblleft low energy\textquotedblright\
effective Hamiltonian 
\begin{eqnarray}
H^{(n)} &=&\omega _{0}(b^{\dagger }b+c^{\dagger }c)+\omega a^{\dagger }a
\label{normal} \\
&&+\lambda a^{\dagger }(c^{\dagger }+b)+\text{H.c.}.  \notag
\end{eqnarray}

We now apply a Bogoliubov transformation to diagonalize the above quadratic
Hamiltonian (\ref{normal}) \cite{book}. First, we rewrite it 
as 
\begin{equation}
H^{(n)}=\frac{1}{2}\mathrm{U}^{(n)\dagger }M^{(n)}\mathrm{U}^{(n)}-\frac{1}{2%
}\mathrm{tr}\text{ }A^{(n)},  \label{060}
\end{equation}%
where the operator-valued vectors $\mathrm{U}^{(n)}$ and the matrices $%
M^{(n)}$, $A^{(n)}$, $B^{(n)}$ are defined as 
\begin{eqnarray}
\mathrm{U}^{(n)} &=&\left( a,b,c;a^{\dagger },b^{\dagger },c^{\dagger
}\right) ^{T},  \notag \\
M^{(n)} &=&\left( 
\begin{array}{cc}
A^{(n)} & B^{(n)} \\ 
B^{(n)\ast } & A^{(n)\ast }%
\end{array}%
\right) ,  \notag \\
A^{(n)} &=&\left( 
\begin{array}{ccc}
\omega & \lambda & 0 \\ 
\lambda & \omega _{0} & 0 \\ 
0 & 0 & \omega _{0}%
\end{array}%
\right) ,  \notag \\
B^{(n)} &=&\left( 
\begin{array}{ccc}
0 & 0 & \lambda \\ 
0 & 0 & 0 \\ 
\lambda & 0 & 0%
\end{array}%
\right) .
\end{eqnarray}%
According to Ref. \cite{book}, we diagonalize the Hamiltonian (\ref{060}) in
two steps: (i) find a unitary canonical transformation $T^{(n)}$ such that $%
T^{(n)}\hat{\eta}T^{(n)\dagger }\hat{\eta}=1$, $T^{(n)\ast }=\hat{\gamma}%
T^{(n)}\hat{\gamma}$, where%
\begin{equation}
\hat{\eta}=\left( 
\begin{array}{cc}
1 & 0 \\ 
0 & -1%
\end{array}%
\right) ,\text{ \ }\hat{\gamma}=\left( 
\begin{array}{cc}
0 & 1 \\ 
1 & 0%
\end{array}%
\right) ;
\end{equation}%
and (ii) introduce the quasiparticle operators%
\begin{equation*}
\mathrm{V}^{(n)}=T^{(n)}\mathrm{U}^{(n)}=(h_{1},h_{2},h_{3};h_{1}^{\dagger
},h_{2}^{\dagger },h_{3}^{\dagger })^{T}.
\end{equation*}%
Then the Hamiltonian (\ref{060}) is cast into a diagonalized form 
\begin{equation}
H^{(n)}=\sum_{i=1}^{3}\omega _{i}^{(n)}(h_{i}^{\dagger }h_{i}+\frac{1}{2})-%
\frac{1}{2}\mathrm{tr}A^{(n)},
\end{equation}%
and describes the quasiparticle excitations with frequencies $\omega
_{1,2,3}^{(n)}$, which are obtained by diagonalizing $\hat{\eta}M^{(n)}$
with $T^{(n)}$ into 
\begin{eqnarray}
T\hat{\eta}MT^{-1} &=&\Omega =\left( 
\begin{array}{cc}
\omega & 0 \\ 
0 & -\omega%
\end{array}%
\right) ,\text{ \ }  \notag \\
\text{\ }\omega &=&\left( 
\begin{array}{ccc}
\omega _{1} &  &  \\ 
& \omega _{2} &  \\ 
&  & \omega _{3}%
\end{array}%
\right) .
\end{eqnarray}



\begin{figure}[h]
\centerline{\includegraphics[width=6cm,height=3.4cm]{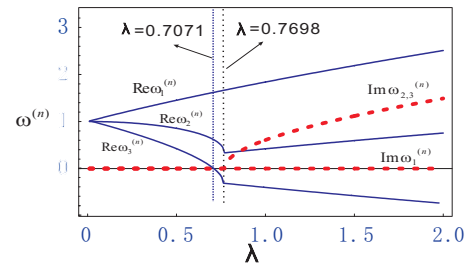}}
\caption{(Color online) The real part (thin solid lines) and imaginary part
(thick dashed lines) of the eigenfrequencies \textit{vs} the coupling
parameter $\protect\lambda $ in units of $\protect\omega _{0}$ in the normal
phase at the resonance case $\protect\omega =\protect\omega _{0}$.}
\label{a1}
\end{figure}

Below, we focus only on the properties of eigen-frequencies in order to
explore the existence of quantum criticality. Since the matrix $M^{(n)}$ is
of $6\times 6$, general analytic results for the diagonalization of $\hat{%
\eta}M^{(n)}$ are difficult to obtain. Nevertheless, we can diagonalize it
numerically to obtain the eigen-frequencies $\omega _{1,2,3}^{(n)}$ \cite%
{reduce}. The related canonical transformation matrix $T^{(n)}$ can also be
obtained numerically. Figure \ref{a1} shows the numerical results for the 
real and imaginary parts of $\omega _{1,2,3}^{(n)}$. For simplicity, we
illustrate the resonant case $\omega =\omega _{0}$ $=$ $1$ here. Certainly,
the non-resonance cases can also be studied numerically, with similar
features being revealed. As seen from Fig. \ref{a1}, when $\lambda >0.7698$,
the imaginary parts of two eigen-frequencies are non-zero. This means that
the corresponding eigen-frequencies are complex and thus the eigen state is
unstable and physically impossible. 
But it is inappropriate to consider naively that $\lambda =0.7698$ as the
critical point. Since the eigenvalue $\omega _{3}^{(n)}$ is negative in the
range of $\lambda \in (0.7071,0.7698)$, a negative eigen-frequency of the
boson-mode is not allowed physically either. Therefore, in the resonance
case $\omega =\omega _{0}$ ($=1$), a real critical point is located at $%
\lambda _{c}^{(n)}=0.7071=\sqrt{2}/2$. In addition, for a general case, the
critical point is found to be $\lambda _{c}^{(n)}=\sqrt{\omega \omega _{0}/2}
$.

\section{Super-radiant phase}

In order to describe excitations in the parameter region above the critical
point, we now incorporate the fact that both the field and the atomic
collective excitations acquire macroscopic occupations, namely, the above
approximation to neglect the number of excitations over $N$ is no longer
valid \cite{emary}. To this aspect, the introduced collective operators $%
B^{\dagger }$ and $C^{\dagger }$ in Eq. (\ref{collective}) should be
expressed approximately as \cite{liusun} 
\begin{eqnarray}
B^{\dagger } &=&b^{\dagger }\sqrt{1-\frac{b^{\dagger }b}{N}},\ \ 
\label{hptrans} \\
C^{\dagger } &=&c^{\dagger }\sqrt{1-\frac{c^{\dagger }c}{N}},  \notag
\end{eqnarray}%
%
%
%
%
%
%
%
%
%
%
%
%
in terms of \ the Bose operators $b^{\dagger },b,c^{\dagger }$ and $c$. This
transformation maps the original light-atoms system to a coupling three-mode
bosonic system with the Hamiltonian%
\begin{eqnarray}
H^{(s)} &=&\omega _{0}(b^{\dagger }b+c^{\dagger }c)+\omega a^{\dagger
}a+\lambda a^{\dagger }  \label{super} \\
&&\times \left( c^{\dagger }\sqrt{1-\frac{c^{\dagger }c}{N}}+\sqrt{1-\frac{%
b^{\dagger }b}{N}}b\right) +H.c..  \notag
\end{eqnarray}

For the present super-radiant phase, 
the bosonic modes may be displaced in the following way: 
\begin{eqnarray}
a^{\dagger } &\rightarrow &d^{\dagger }+\alpha ^{\ast };\text{ }  \notag \\
b^{\dagger } &\rightarrow &e^{\dagger }-\beta ^{\ast };\text{ }  \notag \\
c^{\dagger } &\rightarrow &f^{\dagger }-\gamma ^{\ast },  \label{030}
\end{eqnarray}%
where $\alpha $, $\beta $ and $\gamma $ are generally \textit{complex}
parameters in the order of $\emph{O}(\sqrt{N})$ \cite{emary} to be
determined later. This is equivalent to assume that all modes behave as the
nonzero, macroscopic mean fields above $\lambda _{c}^{(n)}$.

Keeping the terms up to the order of $\emph{O}(N^{0})$, the Hamiltonian (\ref%
{super}) becomes 
\begin{eqnarray}
H^{(s)} &=&\left\{ \lambda \sqrt{\frac{k_{\mathrm{f}}}{N}}\left[ d^{\dagger
}f^{\dagger }+\frac{(\alpha ^{\ast }f^{\dagger }-\gamma ^{\ast }d^{\dagger })%
}{2k_{\mathrm{f}}}(\gamma f^{\dagger }+\gamma ^{\ast }f)\right] \right. 
\notag \\
&&\left. +\lambda \sqrt{\frac{k_{\mathrm{e}}}{N}}\left[ de^{\dagger }+\frac{%
(\alpha e^{\dagger }-\beta ^{\ast }d)}{2k_{\mathrm{e}}}(\beta e^{\dagger
}+\beta ^{\ast }e)\right] +h.c.\right\}  \notag \\
&&+\omega d^{\dagger }d+\omega _{e}e^{\dagger }e+\omega _{f}f^{\dagger }f+%
\mathrm{c}_{\mathrm{0}}^{(s)},  \label{Hamil}
\end{eqnarray}%
where the constant term 
\begin{eqnarray}
\mathrm{c}_{\mathrm{0}}^{(s)} &=&\omega \left\vert \alpha \right\vert
^{2}+\omega _{0}(\left\vert \beta \right\vert ^{2}+\left\vert \gamma
\right\vert ^{2})  \notag \\
&&-\frac{2\lambda \alpha }{\sqrt{N}}(\beta ^{\ast }\sqrt{k_{\mathrm{e}}}%
+\gamma \sqrt{k_{\mathrm{f}}})
\end{eqnarray}%
will substantially contribute to the ground state energy at critical point;
the renormalized frequencies%
\begin{eqnarray*}
\omega _{e} &=&\omega _{0}+\lambda \alpha ^{\ast }\beta /\sqrt{Nk_{\mathrm{e}%
}}, \\
\omega _{f} &=&\omega _{0}+\lambda \alpha \gamma /\sqrt{Nk_{\mathrm{e}}},
\end{eqnarray*}%
with $k_{\mathrm{e}}=N-\left\vert \beta \right\vert ^{2}$ and $k_{\mathrm{f}%
}=N-\left\vert \gamma \right\vert ^{2}$%
%
%
%
. In the derivation of Eq. (\ref{Hamil}), the terms being linear in the
bosonic operators are eliminated by choosing the appropriate displacements $%
\alpha $, $\beta $ and $\gamma $ in the following four cases: 
\begin{eqnarray}
&&\alpha ^{(1,2,3,4)}=\frac{2\mathrm{e}^{i\phi }\lambda }{\omega }\sqrt{%
\frac{X_{+}X_{-}}{N}},  \label{alpha} \\
&&\left\{ 
\begin{array}{l}
\beta ^{(j)}=\mathrm{e}^{i\phi }\sqrt{X_{\mp }},\ \ \ \text{(}j=1,3\text{)}
\\ 
\gamma ^{(j)}=\mathrm{e}^{-i\phi }\sqrt{X_{\mp }},\text{\ (}j=1,3\text{)},%
\end{array}%
\right.  \notag \\
\text{or \ } &&\left\{ 
\begin{array}{l}
\beta ^{(j)}=\mathrm{e}^{i\phi }\sqrt{X_{\pm }},\ \ \ \text{(}j=2,4\text{)},
\\ 
\gamma ^{(j)}=\mathrm{e}^{-i\phi }\sqrt{X_{\mp }},\text{ (}j=2,4\text{)},%
\end{array}%
\right.  \notag
\end{eqnarray}%
%
%
%
%
%
%
%
%
%
%
%
%
%
%
%
%
%
%
%
where $X_{\pm }=\frac{N}{2}(1\pm \frac{\omega \omega _{0}}{2\lambda ^{2}})$,
and $\phi $ is an arbitrary real number relating to the phases of
displacements. In fact, we see from the form of $\alpha ^{(j)}$ in Eq. (\ref%
{alpha}) that only when 
\begin{equation}
4\lambda ^{4}-\omega ^{2}\omega _{0}^{2}\geq 0,
\end{equation}%
$\alpha ^{(j)}$\ can be physically meaningful. Thus in the following
discussions, it is required that 
\begin{equation*}
\lambda \geqslant \sqrt{\frac{\omega \omega _{0}}{2}}\ (=\lambda _{c}^{(n)}).
\end{equation*}%
It is interesting to note that this threshed is just the critical point
determined in the normal phase case.

Since $H^{(s)}(\phi )$ in Eq. (\ref{Hamil}) can be transferred to a $\phi $%
-independent Hamiltonian $H^{(s)}(\phi \equiv 0)$ through a unitary
transformation 
\begin{equation*}
U(\phi )=e^{i\phi (d^{\dagger }d+e^{\dagger }e-f^{\dagger }f)},
\end{equation*}%
we need only to 
look into the spectra of $H^{(s)}(0)$ in the four cases specified by Eq. (%
\ref{alpha}), respectively. Because $H^{(s)}(0)$ is quadratic in each case,\
which is diagonalized by using the same method presented above as 
\begin{eqnarray}
H^{(j)} &=&\sum_{i=1}^{3}\omega _{i}^{(j)}\left( e_{i}^{(j)\dagger
}e_{i}^{(j)}+\frac{1}{2}\right)  \notag \\
&&-\frac{1}{2}\mathrm{tr}\text{ }A^{(j)}+\mathrm{c}_{\mathrm{0}}^{(j)},\ 
\end{eqnarray}%
for the four cases $j=1,...,4$. Here, the quasi-particle excitation is
described by the boson vector operators 
\begin{eqnarray*}
\mathrm{e}^{(j)} &=&(e_{1}^{(j)},e_{2}^{(j)},e_{3}^{(j)};e_{1}^{(j)\dagger
},e_{2}^{(j)\dagger },e_{3}^{(j)\dagger })^{\mathrm{T}} \\
&=&T^{(j)}\left( d^{(j)},e^{(j)},f^{(j)};d^{(j)\dagger },e^{(j)\dagger
},f^{(j)\dagger }\right) ^{\mathrm{T}}
\end{eqnarray*}%
in the $j$-th phase, where 
\begin{equation*}
d^{(j)}=a-\left\vert \alpha ^{(j)}\right\vert ,\text{ }e^{(j)}=b+\left\vert
\beta ^{(j)}\right\vert ,\text{ }f^{(j)}=c+\left\vert \gamma
^{(j)}\right\vert
\end{equation*}%
according to Eq. (\ref{030}). $T^{(j)}$ is still the introduced unitary
transformation to diagonalize%
\begin{equation*}
\hat{\eta}M^{(j)}=\left( 
\begin{array}{cc}
A^{(j)} & B^{(j)} \\ 
-B^{(j)} & -A^{(j)}%
\end{array}%
\right) ,
\end{equation*}%
where 
\begin{eqnarray*}
A^{(1,3)} &=&\left( 
\begin{array}{ccc}
\omega & A_{\mp } & C_{\mp } \\ 
A_{\mp } & \omega _{\mp } & 0 \\ 
C_{\mp } & 0 & \omega _{\mp }%
\end{array}%
\right) , \\
B^{(1,3)} &=&\left( 
\begin{array}{ccc}
0 & C_{\mp } & A_{\mp } \\ 
C_{\mp } & B_{\mp } & 0 \\ 
A_{\mp } & 0 & B_{\mp }%
\end{array}%
\right) ,
\end{eqnarray*}%
\begin{eqnarray*}
A^{(2,4)} &=&\left( 
\begin{array}{ccc}
\omega & A_{\pm } & C_{\mp } \\ 
A_{\pm } & \omega _{\pm } & 0 \\ 
C_{\mp } & 0 & \omega _{\mp }%
\end{array}%
\right) , \\
B^{(2,4)} &=&\left( 
\begin{array}{ccc}
0 & C_{\pm } & A_{\mp } \\ 
C_{\pm } & B_{\pm } & 0 \\ 
A_{\mp } & 0 & B_{\mp }%
\end{array}%
\right) ,
\end{eqnarray*}%
with%
\begin{eqnarray*}
\omega _{\pm } &:&=\omega _{0}+4\lambda ^{2}X_{\pm }/N\omega , \\
B_{\pm } &:&=2\lambda ^{2}X_{\pm }/N\omega , \\
A_{\pm } &:&=\lambda (X_{\mp }-X_{\pm }/2)/\sqrt{NX_{\mp }}, \\
C_{\pm } &:&=-\lambda X_{\pm }/2\sqrt{NX_{\mp }}.
\end{eqnarray*}%
Clearly, $\omega _{i}^{(j)}$ ($i=1,2,3$) is the $i$-th eigenfrequency for
the Hamiltonian $H^{(j)}$. 
Note that the canonical transformation matrix $T^{(j)}$ can be obtained
numerically in the numerical diagonalization of $\hat{\eta}M^{(j)}$.

\begin{figure}[h]
\centerline{\includegraphics[width=8.5cm,height=4.2cm]{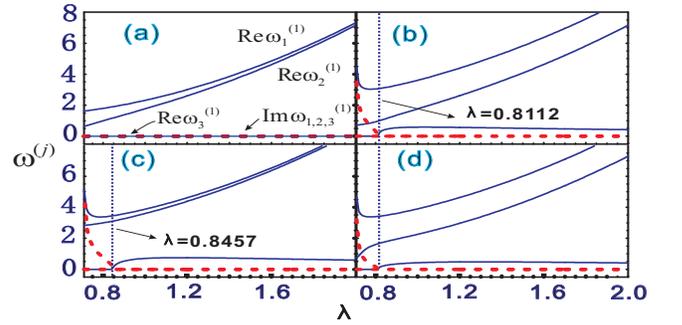}}
\caption{(Color online) The real part (solid lines) and imaginary part
(dashed lines) of the eigenfrequencies \textit{vs} the coupling parameter $%
\protect\lambda $ in units of $\protect\omega _{0}$ in the super-radiant
phase at the resonance case $\protect\omega =\protect\omega _{0}$, where
(a,b,c,d) correspond respectively to the cases (1,2,3,4) specified in the
text.}
\label{a2}
\end{figure}

The numerical results of eigen-frequencies $\omega _{1,2,3}^{(j)}$ \textit{%
vs.} the coupling parameter $\lambda $ are plotted in Fig. \ref{a2}. The
curves for both the real and imaginary parts of eigenfrequencies of $H^{(1)}$
in the resonant case are shown Fig. \ref{a2}(a). It is found 
that the eigenfrequencies are physically reasonable when $\lambda >0.7071$
since the imaginary parts of all the eigenfrequencies are zero. This means a
novel \textquotedblleft quantum phase" emerges above the critical point%
\begin{equation*}
\lambda _{c}^{(1)}=\sqrt{\frac{\omega \omega _{0}}{2}}=0.7071.
\end{equation*}%
It is seen from Fig. \ref{a2}(a) that the eigenfrequency $\omega _{3}^{(1)}$
is always zero above $\lambda _{c}^{(1)}$, which implies that $H^{(1)}$ is
reduced to have two independent boson modes. It is remarkable that the
critical point is just the same one as that determined in the normal phase $%
\lambda _{c}^{(1)}=\lambda _{c}^{(n)}=\lambda _{c}$, demonstrating the
consistency of our analysis. 
%
In Fig. \ref{a2}(b) [Fig. \ref{a2}(d)], the numerical results for the real
and imaginary parts of three eigenfrequencies of $H^{(2)}$ [$H^{(4)}$] in
the resonant case. As seen from Fig. \ref{a2}(b) [Fig. \ref{a2}(d)], only
when 
\begin{equation*}
\lambda >\lambda _{c}^{(2)}=0.8112\ (\lambda _{c}^{(4)}=0.8112),
\end{equation*}%
another possible \textquotedblleft quantum phase" may appear as the
imaginary parts of all the eigenfrequencies are zero. 
While for $H^{(3)}$, as seen from Fig. \ref{a2}(c), 
only when $\lambda >0.8457=\lambda _{c}^{(3)}$, the imaginary part of the
all eigenfrequencies are zero, indicating a possible \textquotedblleft
quantum phase.\textquotedblright\ 
\begin{figure}[h]
\centerline{\includegraphics[width=8cm,height=3.0cm]{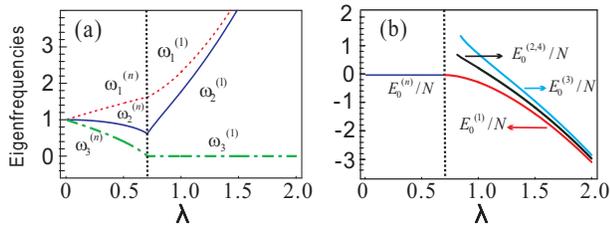}}
\caption{(Color online) (a) The eigenfrequencies for normal phase and the
first super-radiant phase. (b) The ground state energy densities ($N=10^{6}$%
) for the normal phase and the super-radiant phases at the resonance case $%
\protect\omega =\protect\omega _{0}=1$.}
\label{a3}
\end{figure}

In Fig. \ref{a3}(a), we plot together the eigenfrequencies vs $\lambda $ for
the normal phase and first super-radiant phase. The eigenfrequencies $\omega
_{i}^{(n)}$ and $\omega _{i}^{(1)}$ (for $i=1,2,3$) are continuous at the
critical point, respectively. Comparing with the results in the spatially
independent Dicke model \cite{emary}, our numerical studies show clearly
that the excitation energy $\omega _{3}^{(n)}$ in the normal phase vanishes
as $\left\vert \lambda -\lambda _{c}\right\vert ^{zv}$ and the
characteristic length scale $l_{3}=1/\sqrt{\omega _{3}^{(n)}}$ diverges as $%
\left\vert \lambda -\lambda _{c}\right\vert ^{-v}$ at the quantum transition
point $\lambda _{c}$, with the exponents given by $v=1/2$, $z=2$ on
resonance; however, it is interesting to note that no critical exponents for 
$\omega _{3}^{(1)}$ in the super-radiant phase can be specified since $%
\omega _{3}^{(1)}\equiv 0$. Meanwhile, for the ground state, $\left\langle
a^{\dagger }a\right\rangle _{G}/N=0$ below $\lambda _{c}$, while%
\begin{equation*}
\frac{\left\langle a^{\dagger }a\right\rangle _{G}}{N}=\left\vert \alpha
^{(1)}\right\vert ^{2}/N\propto (\lambda -\lambda _{c})
\end{equation*}%
above $\lambda _{c}$, i.e., the field is macroscopically occupied. So $%
\alpha ^{(1)}$ may be understood as a kind of order parameter of the
super-radiant phase, whose critical exponent is $1/2$ above $\lambda _{c}$.
In addition, Fig. 3(b) presents the ground-state energy densities as a
function of coupling for all the possible phases. Clearly, the ground state
energy densities of the first super-radiant phase is always the lowest one
above $\lambda _{c}$, while the other three approach to it in the large $%
\lambda $ limit; moreover, it connects continuously with that of the normal
phase but possesses a discontinuity in its second derivative at $\lambda
_{c} $ through a detailed numerical analysis. From this viewpoint, together
with the fact that the same critical point is determined from both sides of
the normal phase and the first one, it is most likely that only the first
super-radiant phase is a real physical one.

\section{Remarks and conclusions}

Before concluding this paper, we wish to remark briefly on the origin of the
occurred QPT in the present work. From Fig. 3(a), it is clear that the
energy level of the first excited state of the system $(\omega
_{3}^{(n)}+E_{0}^{(n)})$ touches the ground state energy level $E_{0}^{(n)}$
(or $E_{0}^{(1)}$) at the critical point. Obviously, it is this level
touching that accounts for the emergence of the QPT and the corresponding
quantum criticality in the present generalized Dicke model. It is also
remarked that the $A^{2}$ terms (where $A$ is the vector potential) has been
neglected here, as done in several previous works \cite{emary,Reslen05,Vidal}%
, while the absence of $A^{2}$ terms \cite{Rzazewski1,Rzazewski2} seems to
be crucial for the observed quantum phase transition in the present model,
namely, the presence of $A^{2}$ terms in the model Hamiltonian leads to
vanishing of the criticality. 
%

Although the effect of non-RWA terms may normally be negligibly small, the
present work (also see Ref. \cite{emary}) illustrates that it plays a
meaningful role when the atomic number $N$ is large, e.g., the criticality
differs from that with the RWA. On the other hand, for actual atoms that may
not be pure two-level ones, other atomic transitions may occur and spoil the
present model before the non-RWA terms become important. Nevertheless, the
present study is still theoretically interesting and valuable, particularly
relevant to some atomic systems (or artificial and atomic-like ones) wherein
the energy spacing of any other transitions is much larger than that of the
considered two levels (or other transitions do not exist). For example, for
a Dicke-like model consisting of many 1/2 spins coupled to single mode
bosonic field (by electrical dipole coupling-like type), other transitions
do not exist in the spin systems. Then the counter rotating terms play an
important role the when the coupling parameter is close to the critical
value.

In conclusion, based on a generalized Dicke model, we have investigated
theoretically the quantum criticality of an extended atomic ensemble with a
larger spacial dimension comparable to the optical wavelength of a quantized
light field. A useful formalism is developed to study numerically
eigenfrequencies of the system in different quantum phases. Comparing with
the critical phenomenon around the critical point $\widetilde{\lambda }_{c}=%
\sqrt{\omega \omega _{0}}/2$ for atomic ensemble of small dimension~\cite%
{emary}, a rather different quantum criticality is revealed around the
transition point ($\lambda _{c}=\sqrt{\omega \omega _{0}/2}=\sqrt{2}{%
\widetilde{\lambda }}_{c}$) from the normal phase to the super-radiant
phase. 

This work was supported by the NSFC with grant Nos. 90203018, 10474104,
60433050, 10447133, 10574133, \& 10429401, the NFRP of China with funding
Nos. 2001CB309310 and 2005CB724508, the RGC grant of Hong Kong
(HKU7045/05P), and the URC fund of HKU.


\begin{thebibliography}{99}
\bibitem{reduce} Actually, for the particlular $6\times 6$ dimensional
matrix $\hat{\eta}M^{(n)}$ in this noraml phase, we have a simple method to
diagonize it by reducing it into two $3\times 3$ dimensional sub-matrices
(not presented here).
\end{thebibliography}
\end{document}